\documentclass[conference]{IEEEtran}
\IEEEoverridecommandlockouts
\pdfoutput=1 
\usepackage[utf8]{inputenc}
\usepackage[T1]{fontenc}
\usepackage{amsmath,amssymb,amsfonts}
\usepackage{algorithmic}
\usepackage[braket, qm]{qcircuit}
\usepackage{graphicx}
\usepackage{textcomp}
\usepackage[table,xcdraw]{xcolor}
\usepackage{xcolor}
\usepackage{multirow}
\usepackage{tabularx,ragged2e,booktabs,caption}
\usepackage{biblatex} 
\usepackage{enumitem}
\usepackage{pifont}
\def\BibTeX{{\rm B\kern-.05em{\sc i\kern-.025em b}\kern-.08em
    T\kern-.1667em\lower.7ex\hbox{E}\kern-.125emX}}
\usepackage{xspace}
\usepackage{flushend}
\usepackage{algorithm2e}
\begin{document}
\RestyleAlgo{ruled}
\title{Hybrid Gate-Pulse Model for Variational Quantum Algorithms
\vspace{-7pt}
}
\renewcommand{\bibfont}{\normalfont\small}
\newcommand{\red}[1]{\textcolor{red}{#1}}
\newcommand{\zx}[1]{\textcolor{cyan}{(ZX: #1)}}
\newcommand{\jlc}[1]{\textcolor{violet}{JLC: #1}}
\newcommand{\hzc}[1]{\textcolor{blue}{Zichang: #1}}

\author{
    \IEEEauthorblockN{Zhiding Liang\IEEEauthorrefmark{1} \ \
    Zhixin Song\IEEEauthorrefmark{2} \ \
    Jinglei Cheng\IEEEauthorrefmark{3} \ \
    Zichang He\IEEEauthorrefmark{4} \ \
    Ji Liu\IEEEauthorrefmark{5} \ \
    Hanrui Wang\IEEEauthorrefmark{6}\\
    Ruiyang Qin\IEEEauthorrefmark{1}\ \
    Yiru Wang\IEEEauthorrefmark{3} \ \
    Song Han\IEEEauthorrefmark{6} \ \
    Xuehai Qian\IEEEauthorrefmark{3} \ \
    Yiyu Shi\IEEEauthorrefmark{1}
    }
    \IEEEauthorblockA{\IEEEauthorrefmark{1}University of Notre Dame \
    \IEEEauthorrefmark{2}Georgia Institute of Technology \ 
    \IEEEauthorrefmark{3}Purdue University \ 
    }
    \IEEEauthorblockA{
    \IEEEauthorrefmark{4}University of California, Santa Barbara \ 
    \IEEEauthorrefmark{5}Argonne National Laboratory \ 
    }
    \IEEEauthorblockA{
    \IEEEauthorrefmark{6}Massachusetts Institute of Technology \ 
    }
    }

\maketitle

\begin{abstract}
Current quantum programs are mostly synthesized and compiled on the gate-level, where quantum circuits are composed of quantum gates.
The gate-level workflow, however, introduces significant redundancy when quantum gates are eventually transformed into control signals and applied on quantum devices.
For superconducting quantum computers, the control signals are microwave pulses.
Therefore, pulse-level optimization has gained more attention from researchers due to their advantages in terms of circuit duration.
Recent works, however, are limited by their poor scalability brought by the large parameter space of control signals.
In addition, the lack of gate-level ``knowledge'' also affects the performance of pure pulse-level frameworks.
We present a hybrid gate-pulse model that can mitigate these problems.
We propose to use gate-level compilation and optimization for ``fixed'' part of the quantum circuits and to use pulse-level methods for problem-agnostic parts.
Experimental results demonstrate the efficiency of the proposed framework in discrete optimization tasks.
We achieve a performance boost at most 8\% with 60\% shorter pulse duration in the problem-agnostic layer.

\end{abstract}


\section{Introduction}
\label{sec:intro}

Quantum Computing (QC) is an emerging technique with the potential to achieve exponential acceleration over classical computation.
QC has promised theoretical speedups for problems including integer factoring~\cite{shor1999polynomial} and unstructured database search~\cite{grover1996fast}. 
However, these algorithms are designed for fault-tolerant quantum computers. 
In the current Noisy Intermediate Scale Quantum (NISQ)~\cite{preskill2018quantum} era, quantum computers with hundreds of qubits are available, but hardware noise still limits the depth of quantum circuits that can be executed reliably. 
Variational Quantum Algorithms (VQAs)~\cite{cerezo2021variational} are among the most promising tasks to demonstrate practical quantum advantages on current noisy quantum computers. These hybrid algorithms utilize Parametrized Quantum Circuits (PQC) with external parameters (e.g., angle $\theta$ in rotation gates such as $R_y(\theta)$) as
a degree of freedom to explore the Hilbert space. 
Then, one can use a classical computer to optimize these parameters with respect to problem-tailored cost functions $C(\vec{\theta})$.
VQAs have been adopted to explore the opportunities of using quantum resources to speed up computations in quantum chemistry~\cite{peruzzo2014variational, o2016scalable, nam2020ground}, combinatorial optimization~\cite{farhi2014quantum, harrigan2021quantum}, and machine learning~\cite{romero2017quantum, zoufal2019quantum}.

\begin{figure}[t]
\centering
\includegraphics[width=1\linewidth]{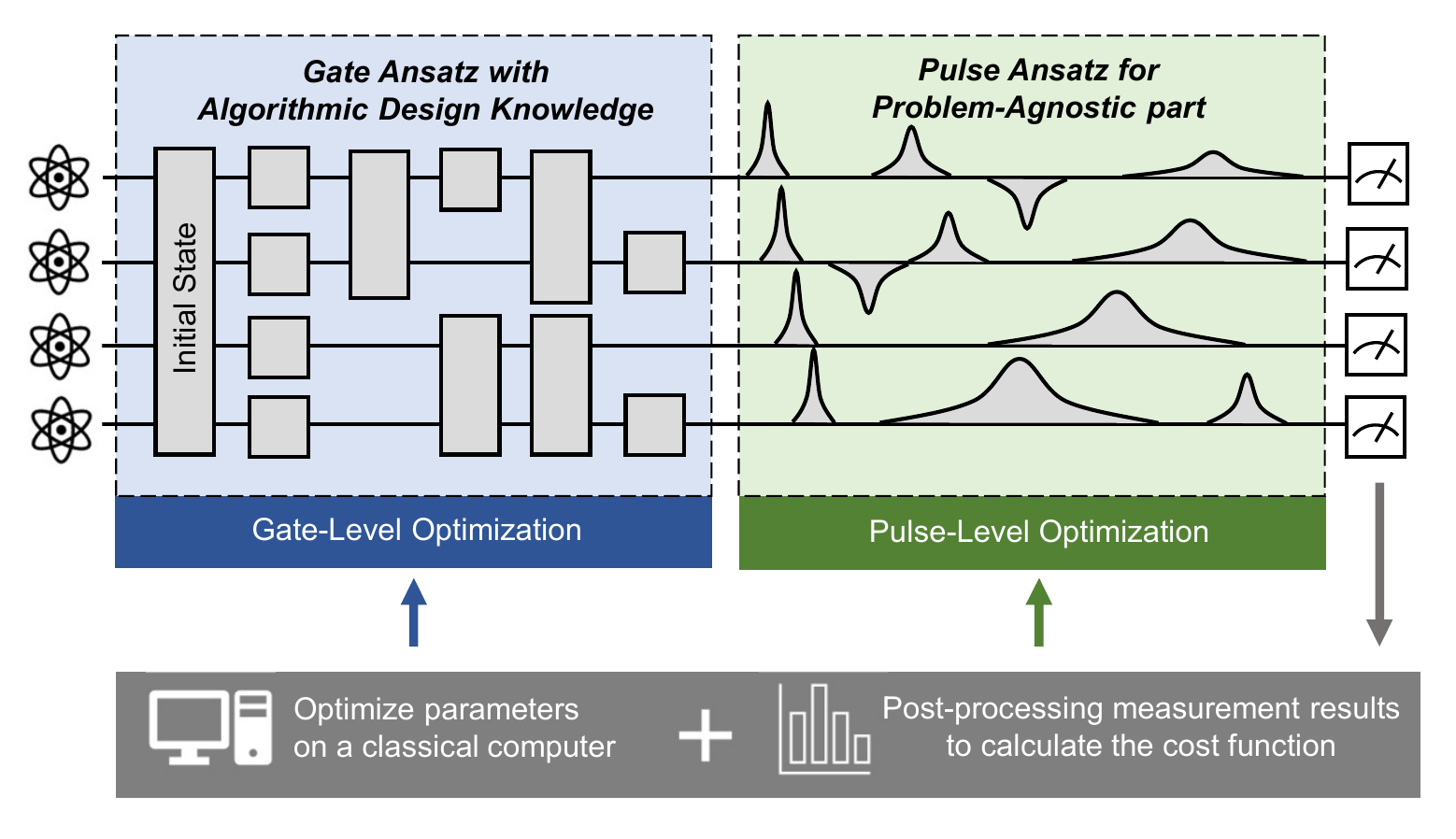}
\caption{Overview of our hybrid optimization framework for VQAs. Here we keep the circuit for initial state preparation in gate-level, since it generally requires high accuracy and gate-level operations are well calibrated. However, whether one can efficiently encode information directly on the pulse-level is an interesting open question.}
\vspace{-5mm}
\label{fig:teaser}
\end{figure}

\begin{figure*}[t]
\includegraphics[width=\linewidth]{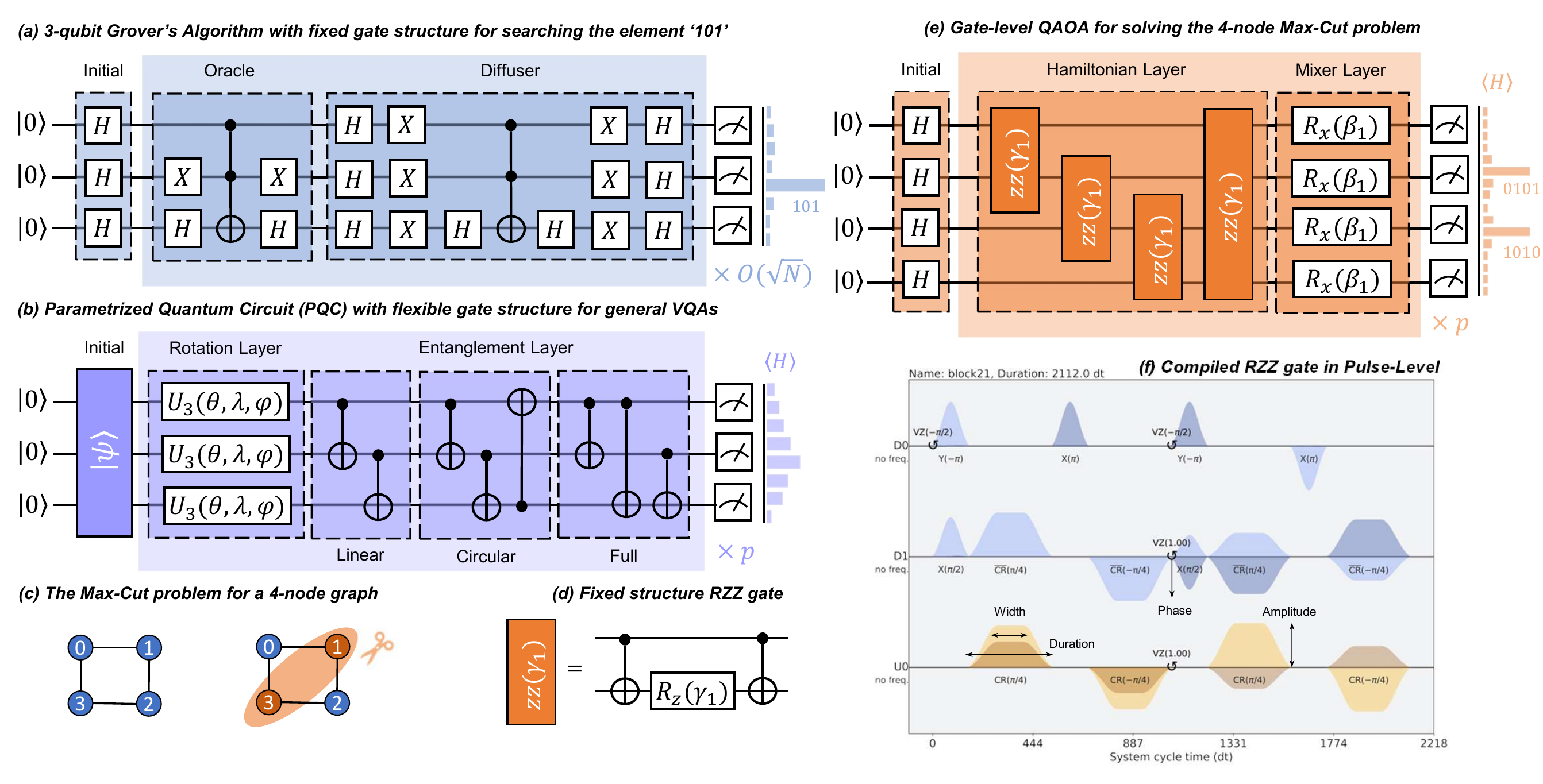}
\caption{(a) Fixed v.s. (b) Flexible algorithm design. Notice the entangling layer consists of CX gates can be customized. 
(c) Visualization of the Max-Cut for a 4-node graph; (d) A fixed design in QAOA due to problem encoding in the Hamiltonian layer; (e) The gate-level visualization of QAOA ansatz;
(f) Pulse-level realization of the RZZ gate in (d). The pulse parameters include amplitude, width, phase, frequency, and duration. 
The Cross-resonance (CR) gate is the most native 2-qubit operation to superconducting qubits. 
All the CX gates in entanglement layer of (b) has to be compiled into CR plus other pulse operations, which introduces redundancy as part of the compilation.
``D'' means the drive channels and ``U'' represents the control channels.
}
\vspace{-5mm}
\label{fig:pulsegraph}
\end{figure*}

Typically, researchers use a quantum circuit, filled with gate-level operations, as a high-level abstraction layer to design and implement quantum algorithms. 
For long-term quantum algorithms such as Grover's algorithm designed for unstructured search, each part of the circuit is constructed with specific ``knowledge'' about the algorithm, which results in less flexibility in the circuit architecture.
On the contrary, VQAs are heuristic algorithms and usually use a problem-agnostic black-box circuit architecture. We notice one special case for VQA is the unitary coupled-cluster (UCC) ansatz~\cite{taube2006new} inspired by quantum chemistry theory. Since it is problem-aware, it also maintains a fixed circuit architecture.
We illustrate such a difference of algorithm design in Fig.~\ref{fig:pulsegraph}a-b. 
However, both of them involve quantum operations that are not native to the hardware. 
To run these gate-level algorithms on real-world NISQ machines, we need to compile them into a lower-level abstraction layer of control signals such as microwave pulses for IBM's superconducting quantum processors or laser operations for Quantinuum's ion-trap quantum computers. 



In the past few years, we have seen multiple works on designing VQAs with quantum pulses~\cite{meitei2021gate, liang2022variational, magann2021pulses}, considering a lower-level abstraction layer for quantum algorithms. 
The major advantage of this approach is the reduction of pulse duration and less compilation overhead since each pulse is directly implementable on the hardware. 
However, this abstraction layer makes the optimization problem in VQAs more challenging. 
It also leads to poor scalability due to an increasing search space dimension (each physical pulse involves more parameters than a quantum gate). 
Consequently, pulse-level VQAs introduces higher training costs than their gate-based counterparts. Many methods have already been proposed to reduce training costs for gate-level VQAs~\cite{hu2022quantum, wang2022qoc, wang2022quantumnas, wang2022quantumnat, liang2021can, jiang2021co, qi2022classical}. One recent relevant work for pulse-level VQAs is enabling gradient descent training~\cite{leng2022differentiable}.

In this paper, we propose a hybrid gate-pulse framework to combine the advantages of the two aforementioned abstraction layers. 
Our approach keeps part of the VQA that is based on algorithmic design knowledge at the gate level but directly works on the pulse design when the sub-circuit is problem-agnostic (see Fig.~\ref{fig:teaser}). 
The pulse-level workflow brings us less latency and reduces decoherence noise. 
On the other hand, the gate-level ``knowledge'' helps reduce the parameter space and cut down the training cost. 
We experimentally evaluate the hybrid framework with the Quantum Approximate Optimization Algorithm (QAOA)~\cite{farhi2014quantum} for the Max-Cut problem on four different IBM's quantum computers. 
Specifically, our work has the following contributions: 
\begin{itemize}
  \item We propose the concept of hybrid abstraction layers to the community. 
  The hybrid of gate and pulse models take the advantage of both abstraction layers: algorithm design knowledge and hardware friendly implementation.  
  \item We propose a gate-pulse co-optimization workflow for VQAs.
  We are the first to experimentally realize hybrid gate-pulse optimization for VQAs on NISQ machines. 
  \item Experimental results on four real quantum computers demonstrate that the proposed method outperforms existing methods in terms of algorithmic performance and duration for solving the Max-Cut problem using QAOA.
\end{itemize}

\section{Background}
\label{sec:background}


\textbf{QAOA for the Max-Cut problem:}
In VQAs, a series of parametrized quantum gates $U(\vec{\theta}) = U_p(\theta_p)\cdots U_2(\theta_2)U_1(\theta_1)$ are used to perform unitary transformation on the statevector $|\psi\rangle(\theta) = U(\vec{\theta})\ket{0}^{\otimes n}$, where $n$ is the number of qubits involved in the computation. Each layer of $U_p(\theta_p)$ is usually composed of parametrized single qubit gate $U_3(\theta,\lambda,\varphi)$ and unparametrized entanglement layer $\{CX_{i,j}\}$ among qubit $i$ and $j$ (see Fig.~\ref{fig:pulsegraph}b). 
The computation results are obtained by repeated measurements of the quantum state as a probability distribution $\Pr(|\psi\rangle) = \{| \langle \psi | z\rangle|^2\}_{z=0\cdots00}^{1\cdots11}$. 

QAOA is originally proposed to approxiamtely solve combinatorial optimization problems. 
One of the most well-known problem is Max-Cut, which aims to find a partition of the graph's $G = (V,E)$ vertices into two complementary sets such that the number of edges between these two sets is the largest (Fig.~\ref{fig:pulsegraph}c). 
Unlike the black-box ansatz that are used in some other VQAs, QAOA ansatz (Fig.~\ref{fig:pulsegraph}e) involves an alternating structure inspired by the Quantum Adiabatic Theorem~\cite{born1928beweis}:
\begin{equation}
U(\vec{\beta}, \vec{\gamma}) = U_M(\beta_p)U_P(\gamma_p)\cdots U_M(\beta_1)U_P(\gamma_1),
\end{equation}
where we evolve the quantum state $|\vec{\beta}, \vec{\gamma} \rangle = U(\vec{\beta}, \vec{\gamma}) |+\rangle^{\otimes n}$ according to a Hamiltonian layer $U_P(\gamma_\ell) = e^{-i\gamma_\ell H_P}$ and a mixer layer $U_M(\beta_\ell) = e^{-i\beta_\ell H_M}$. The Hamiltonian $H_P$ encodes the Max-Cut problem such that the ground state is the optimal solution. Then, we can write the optimization as
\begin{equation}
\max C(\vec{\beta}, \vec{\gamma}) = \max  \langle \vec{\beta}, \vec{\gamma}| H_P|\vec{\beta}, \vec{\gamma} \rangle,
\end{equation}
where $C(\vec{\beta}, \vec{\gamma})$ is the cost function. In this paper, we follow the original protocol~\cite{farhi2014quantum} to choose $H_M = X^{\otimes n}$ as the mixer layer and set the initial state as $|+\rangle^{\otimes n}$, a uniform superposition of all the solutions created by a series of Hadamard gate $H$. We choose the approximation ratio (AR) $\alpha = C^*/C_{max}$ to measure how close the approximate solution is to the actual optimum, where $C^*$ is the final value of cost function. 

\textbf{Quantum Pulse Control:}
For superconducting quantum computers, the pulse-level is the lowest layer of the workflow in quantum computing \cite{doria2011optimal, cheng2020accqoc}. 
A pulse is defined as a sequence of analog control signals that are applied to quantum channels of different qubits. 
In IBM's quantum devices, multiple channels are defined.
{\it DriveChannel} is the primary quantum channel that is associated with qubits, while {\it ControlChannel} only exists for multi-qubit operations and is generated by the feedback Hamiltonian from the backend. 
The {\it AcquireChannel} stores the measured data and information. 
The {\it MeasureChannel} generates readout pulses from another resonator, and readout signals are collected and processed after the measurement pulses are applied. 

\section{Motivation}
\label{sec:motivation}
The quantum programs are usually firstly coded in high-level programming languages, and then compiled into assembly languages such as .qasm files.
Since the qubits in current quantum hardware are not fully-connected, some gates in .qasm files can not be performed on a quantum computer. 
The conflicts need to be solved through the insertion of SWAP gates. 
The process is referred to as ``qubit mapping and routing''~\cite{li2019tackling, peham2022optimal,tan2020optimal}.
The mapped .qasm files are then translated into control-pulses and executed on a quantum computer. 
The transpiling process introduces significant redundancy in the aspect of circuit duration. 
Moreover, limited decoherence time in NISQ machines makes it hard to implement large-scale quantum algorithms.
Recently, more researchers have turned their focus from the gate-level abstraction layer to beyond the gate-level abstraction layer. 

Most of the "beyond gate level" works are on the pulse-level abstraction layer.
Existing pulse-level works include applying insights from pulse properties to generate pulse-efficient gate circuit transpilation~\cite{earnest2021pulse, gokhale2020optimized}. 
And~\cite{niu2022effects, ibrahim2022pulse, melo2022pulse} take the advantages of pulse-efficient transpilation to boost the performance of VQAs.
On the other hand,~\cite{liang2022pan, meitei2021gate} directly generate ansatz and implement quantum algorithms on pulse level. 
However, these pulse-level methods lack knowledge from algorithm design, which results in a large parameter space that causes troubles for optimizers.
To address the challenge, we provide our solution on a new abstraction layer. 
We aim to find a good trade-off between the latency and training cost via gate-pulse co-design. 
Our hybrid model is compatible with available gate-level optimization techniques as well as those on pulse-level.

\section{Hybrid Gate-Pulse Model}
\label{sec:hybridmode}
A gate-level VQA with problem-inspired knowledge has the highest latency but the lowest training cost, while a pulse-level model has the lowest latency but the highest training cost. 
As a new abstraction layer for "beyond gate level" endeavors, we propose a hybrid gate-pulse model.
We reduce the redundancy and shorten the duration compared with gate-level framework. 
In addition, it considers gate-level algorithmic design knowledge.
Therefore, it needs less parameters than a pulse-level framework.
Our hybrid gate-pulse model takes a good trade-off between the latency and training cost. We illustrate the workflow to build a hybrid gate-pulse model in Fig.~\ref{fig:overview}:
\begin{itemize}[leftmargin=*]
    \item Step I. We adopt pulse-level optimization to reduce the duration. 
    \item  Step II. We use gate-level optimization methods like gate cancellation, Sabre qubits mapping~\cite{li2019tackling}, etc. 
    \item Step III. We design an error suppression protocol that includes measurement error mitigation algorithms which are also compatible with our model. 
    In this step, general optimization techniques including Conditional Value-at-Risk (CVaR)~\cite{barkoutsos2020improving} can also be applied to further boost the performance of the hybrid model.
\end{itemize}
Since the entire optimization process is conducted in a machine-in-loop way, the framework is resilient to device noises.
The proposed hybrid model is orthogonal to available optimization techniques like dynamic decoupling, noise injection, etc.
These techniques can be applied to our framework. 

\subsection{Basic Model Design and Implementation}
In the proposed model, we keep the deterministic part based on quantum information theory, quantum physics, physical constraints, and domain knowledge associated with the task at the gate level.
For example, in QAOA, we design the Hamiltonian layer at the gate level to maintain the RZZ structure which is based on conventional problem encoding. 
Then, we replace the gates of the problem-agnostic parts with a constructed native-pulse ansatz. 
For example, when dealing with QAOA tasks, we build the mixer layer with the native pulses since the mixer layer is problem-agnostic and hence relatively flexible by design. According to the adiabatic theorem, we only need to prepare an easily accessible ground state $|\psi\rangle_{initial}$ of the mixer Hamiltonian $H_M$ as initialization. Our hybrid model can, in principle, help find novel mixer designs given a fixed initial state.

\subsubsection{Pulse Parameters Setting}
Prior works at the pulse level demonstrate that we can directly tune parameters that cannot be accessed at the gate level. 
Pulse amplitude is one of the most commonly used parameters in the existing literature. 
Both the pulse amplitudes of the {\it DriveChannel} and the {\it ControlChannel} are directly related to the signal's intensity, which determines the drive Hamiltonian. 
Concurrently, it has been demonstrated in existing work that the adjustment of frequency has a significant impact on quantum operations. 
In addition, we can access the phase of classical electronics at the pulse level, which corresponds to the angles of rotation gates at the gate level. 
Therefore, the parametric pulses in our gate-pulse model contain more parameters including frequency, amplitude, and phase. 
Among these parameters, amplitude and frequency are invisible to gate-level users.

\subsubsection{Frequency Tuning for Pulses}
In the gate-pulse model, the boundaries of phase and amplitude are clearly defined.
The absolute value of amplitude cannot exceed 1, and the range of phase is $(0, 2\pi)$. 
In comparison, it is worthwhile to consider how to determine the range of frequency since the change of the power of drive frequency could trigger the frequency collision related to Stark shift~\cite{schuster2005ac}. 
Thus, we consider frequency tuning as the operation against noise from hardware or environment. 
These noises can lead to variances in the accuracy of the actual frequency applied to the machine. 
Therefore, we introduce a flexible method of frequency modulation, which modifies the frequency of each pulse in a more flexible manner than the crude methods that applied to an entire qubit channel and set the range of frequency modulation between -100MHz to 100MHz. 
Our method applies frequency shifts as parameters to parametric pulses, enabling more flexible frequency control for each pulse operation. 
\subsection{Pulse-level Knowledge}
\begin{figure}[t]
\centering
\includegraphics[width=\linewidth]{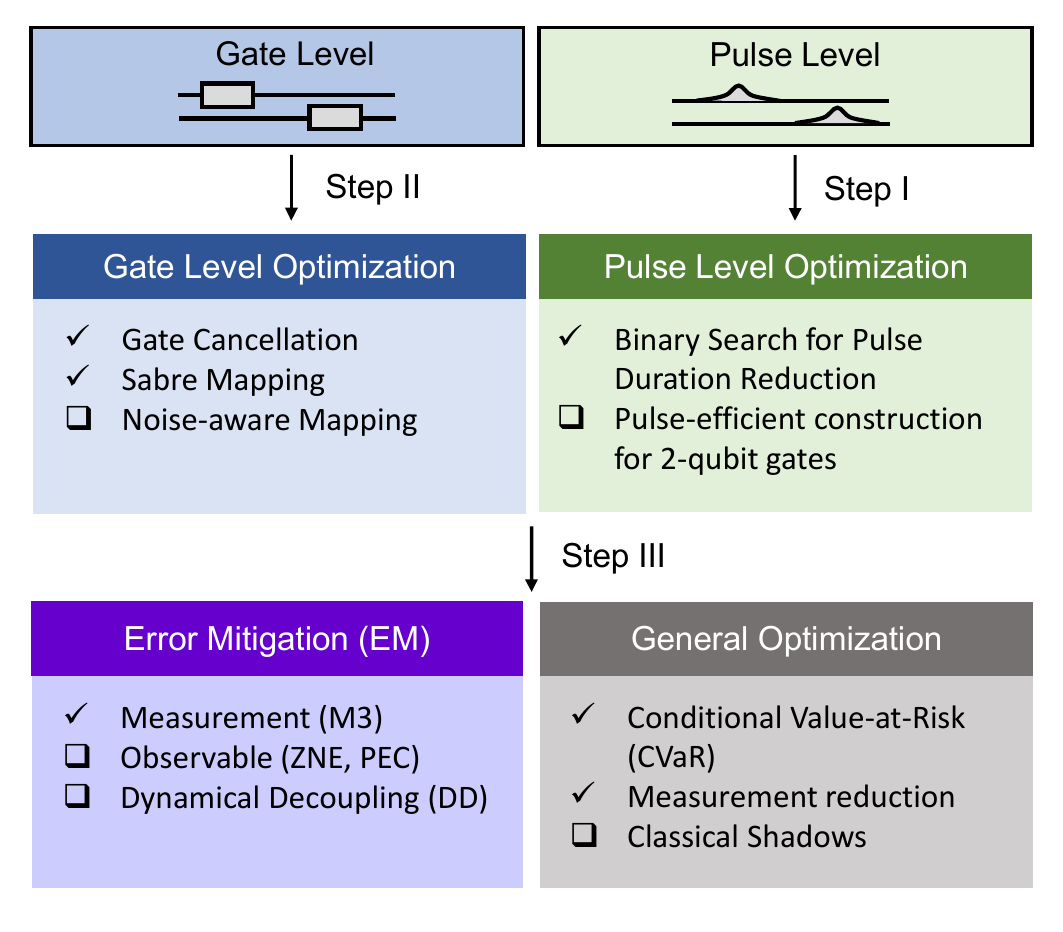}
\caption{The workflow of gate-pulse co-optimization on proposed hybrid gate-pulse model. Step I is pulse-level optimization, followed by gate-level optimization in Step II. Step III aims to design an error suppression protocol for the proposed model, including measurement error mitigation and so on. The methods that mark as selected are used in this project.
}
\vspace{-5mm}
\label{fig:overview}
\end{figure}

The pulse-level optimization is described as Step I in Fig.~\ref{fig:overview}, where we deploy the principles of quantum optimal control (QOC)~\cite{cheng2020accqoc, magann2021pulses}.
We use a binary search algorithm for the pulse duration of the pulse layer in our hybrid gate-pulse model.
In the model, we define the parametric pulse and set its duration to an initial value that is a multiple of 32dt, which is a restriction of the qiskit pulse for Gaussian waveforms. 
Based on the initial value, we use a binary search to search for the minimum duration that is needed for pulses while we maintain the good performance of the model.

\subsection{Gate-level Knowledge}
As discussed in the preceding section, pulse-level models with heavy parameters always exceed the capabilities of the optimizer and can result in a terrible optimization landscape for specific benchmarks. In addition, for certain problems, there exist layers with a fixed structure derived from the knowledge of algorithmic design, such as QAOA, the quantum algorithm for topological data analysis (TDA)~\cite{quantumtda}, the quantum algorithm for partial differentiable equations~\cite{lubasch2020variational}, etc. There are two implementation challenges for these layers at the pulse level: 
1) If the circuit is optimized directly at the pulse level, the final state is unknown unless state tomography is used to determine the final state. For example, the Hamiltonian layer of QAOA fixes the RZZ structure, making it difficult to maintain the structure at the pulse level. This issue could be addressed if the problem is designed with pulse-level encoding, but it will be long-term work with lots of challenges. 
2) Even if we directly transform the gate-level circuit with algorithmic design knowledge to pulse, the calibration could be a problem. Typically, both single-qubit and 2-qubit gates are created by irradiating the qubits with pulses. These corresponding gate and pulse pairs are carefully calibrated and can be carefully controlled in hardware. However, if we control the pulse parameters to form the corresponding circuit, we lose the advantage of careful calibration. Because the pulse parameters given to the hardware will still produce a certain variance due to noise interference and thus cannot achieve precise control.

Therefore, we inherit and retain the knowledge at the gate level in the problem-specific layer of the corresponding problem. In addition, we employ some gate-level optimization techniques to confirm the compatibility of the hybrid gate-pulse model with the gate-level optimization techniques and further optimize the performance of the model for problem-solving as the Step II in Figure \ref{fig:overview}. We used Sabre mapping for better-selecting qubits routing relation, and commutative cancellation to cancel redundant such as self-joint gates via commutation relations~\cite{tan2020optimal} for the benchmarks chosen for this paper. Notably, many gate-level optimizations can be further applied here, but we have selected only a few for demonstration.

\subsection{General Optimization Technique}
In addition to pulse-level and gate-level co-optimization, we can design effective error suppression protocols for the hybrid gate-pulse model, thereby enhancing this model's performance in implementing quantum algorithms on quantum hardware. For example, dynamic decoupling (DD)~\cite{das2021adapt} sequences are incorporated at the algorithmic level to mitigate various idling errors, including dephasing and ZZ crosstalk~\cite{murali2020software}. As for measurement error mitigation, all shots returned from the backend undergo a final postprocessing step. A designed initial calibration program identifies the measurement error process with the help of an algorithm and stores the corresponding calibration data. Finally, calibration data and measurement results are considered in combination at runtime to estimate probability distributions and confidence intervals for the hardware's bitstrings. These two are methods commonly used to design error suppression protocols. And in our work, we applied measurement error mitigation to demonstrate the usefulness of such protocol on the hybrid gate-pulse model. In more detail, we use a matrix-free measurement mitigation (M3) routine~\cite{nation2021scalable}. Instead of forming a full assignment matrix or its inverse, This method operates in a subspace defined by the to-be-corrected noisy input bit strings. Since the number of distinct bit strings can be smaller than the multi-bit Hilbert space's dimensionality, the linearity problem can be solved.

Next, we add methods to the protocol that perform a general performance optimization using the conditional risk value as an aggregation function, allowing quantum algorithms to converge on a better solution quickly. Simply put, in this method, we must define a proportion of the risk value that determines our payout. Get the counts in the final states corresponding to that proportion, then divide this number by the product of the total number of shots and the proportion.



\section{Experiments}
\label{sec:exp}
 
\subsection{Experiment Setups}
\begin{figure}[h]
\centering
\includegraphics[width=1\linewidth]{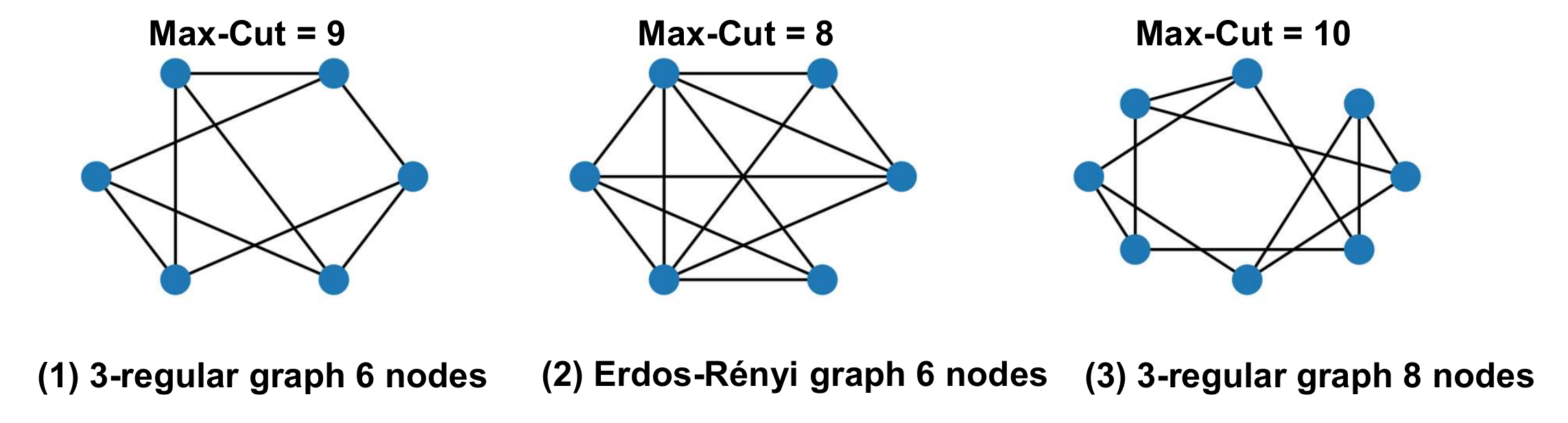}
\caption{Graphs used in the QAOA Max-Cut benchmark. (1)(2)(3) corresponding to task 1, task 2, task 3 below.}
\label{fig:graphs}
\vspace{-3mm}
\end{figure}

Our experiments are conducted on real NISQ machines, including four IBM quantum systems: $ibm\_auckland$, $ibmq\_toronto$, $ibmq\_montreal$, and $ibmq\_guadalupe$. The calibration data is captured in Table.~\ref{table:calibration}. Notice, $ibmq\_toronto$ has the lowest CNOT error rate and we expect the best performance on this backend. $ibm\_auckland$ has the lowest readout error and we expect measurement error mitigation has the least effect on it. We choose level 1 ($p=1$) QAOA for the Max-Cut problem on three-regular six nodes, three-regular eight nodes, and randomized six nodes graphs as our evaluation benchmark as shown in Fig. \ref{fig:graphs}. 
During the optimization procedure, we use the `COBYLA' optimizer and set the maximum iteration to 50. In each iteration, in order to evaluate the cost function $C(\vec{\beta}, \vec{\gamma})$, the hybrid quantum program is executed 1024 times repeatedly for measurement purpose. Furthermore, we fixed the logical to physical qubit mapping and set the CVaR coefficient to 0.3 for fair comparison among experiments. 
 \begin{table}[h]
\small
\centering
 \renewcommand*{\arraystretch}{1}
\setlength{\tabcolsep}{2.4pt}
\begin{tabular}{c c c c c}
 \toprule
 Backends & $auckland$ & $toronto$ &  $guadalupe$ & $montreal$\\
 \hline \hline 
 \# qubit & 27 & 27 & 16 & 27\\
 Pauli-X error & 2.229e-4 & 2.774e-4 & 3.023e-4 & 2.780e-4\\ 
 CNOT error & 1.164e-2 & 9.677e-3 & 1.108e-2 & 1.049e-2\\
 Readout error &  0.011 & 0.031  & 0.025 & 0.015\\
 \hline 
 $T_1$ time (ms) &166.220 & 104.200 & 102.320 & 123.99\\
 $T_2$ time (ms) &145.620 & 120.760 & 102.530 & 95.01\\
 Readout length (ns) &757.333 &  5962.667 & 7111.111 & 5201.778\\
 \toprule
 \end{tabular}
\caption{Calibration data of quantum computers used in this study obtained at the time the experiments were performed.}
\label{table:calibration}
\vspace{-4mm}
\end{table}

 \subsection{Main Results}
 
\begin{figure}[t]
\centering
\includegraphics[width=1\linewidth]{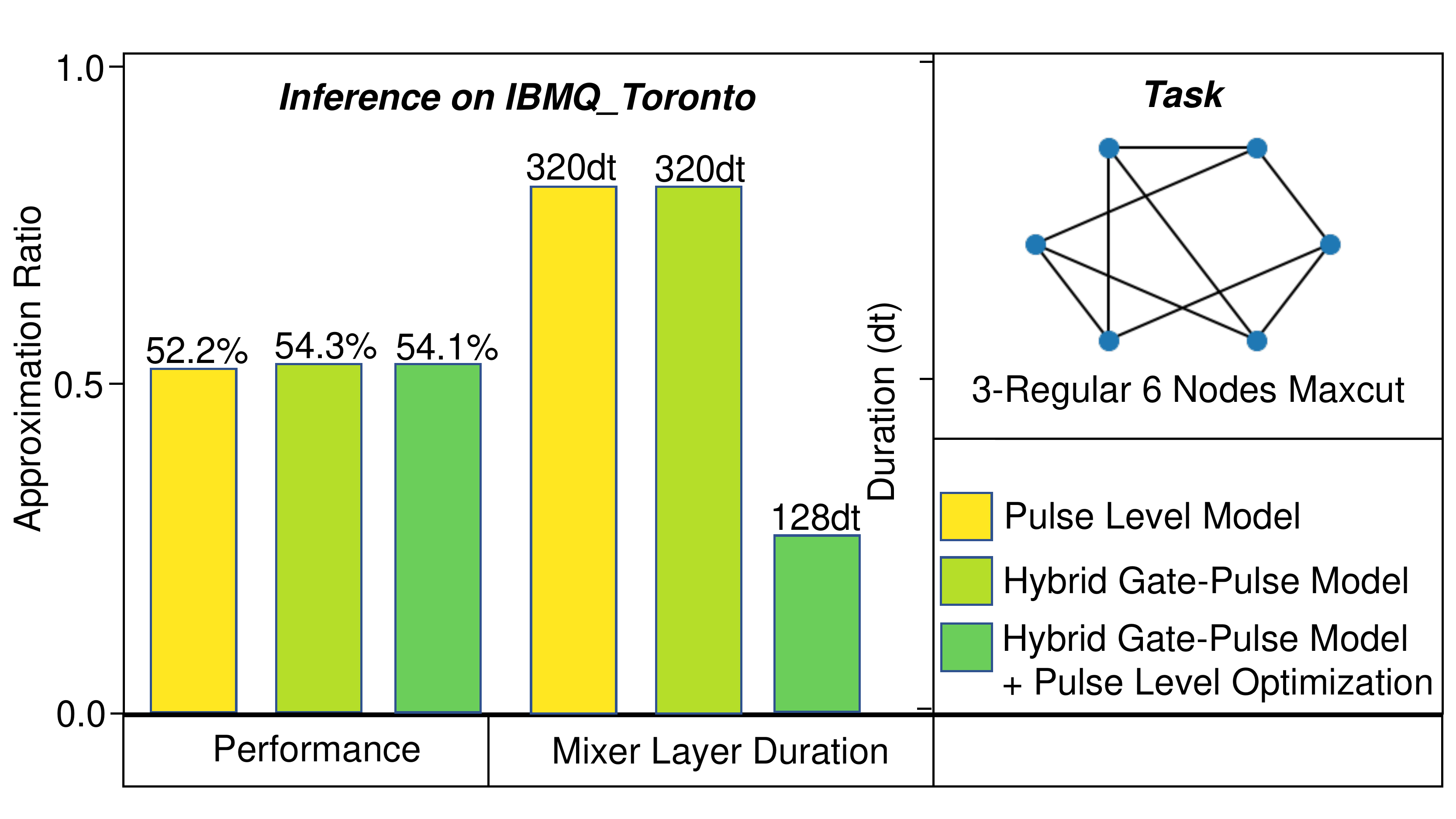}
\caption{Pulse-level model and hybrid gate-pulse model results for QAOA Max-Cut problem with a three-regular six nodes graph on $ibmq\_toronto$, with the pulse duration reduction.}
\label{fig:pulseoptimization}
\vspace{-5mm}
\end{figure}

 \textbf{Pulse-level Optimization: }The results collected from $ibmq\_toronto$ in Fig.~\ref{fig:pulseoptimization} verify the pulse-level optimization techniques are compatible with our proposed model. The binary search optimization method can successfully reduce the 60\% pulse duration of the mixer layer in QAOA with no significant effect on the final approximation ratio. On the other hand, our hybrid model outperforms the pulse-level model with a 2.1\% higher approximation ratio and 4x faster training time to reach convergence. The pulse-level model is initialized from the gate-level circuit (see Fig.~\ref{fig:pulsegraph}) and we gradually lose the fixed structure $Z_iZ_j(\gamma_\ell)$ in the Hamiltonian layer since many pulse parameters (amplitude, frequency, and phase) are changing during optimization. This setup is similar to the VQP approach~\cite{liang2022variational}. Such a loss of knowledge about algorithm design leads to a larger parameter space and hence requires longer convergence time (maximum iteration up to 200). 

 \begin{table*}[t]
 \renewcommand*{\arraystretch}{1}
\setlength{\tabcolsep}{1.4pt}
\small
\centering
\begin{tabular}{c c c c c c c}
\toprule
 Backends & $auckland$ (gate) & $auckland$ (hybrid) & $toronto$ (gate) & $toronto$ (hybrid) & $guadalupe$ (gate) & $guadalupe$ (hybrid)\\ [0.5ex] 
 \hline\hline
 Raw AR & 49.1\% &\textbf{54.2\%}  & 48.8\% & \textbf{54.1\%} & 50.5\% & \textbf{54.5\%} \\ 
 GO AR & 53.3\% &\textbf{55.7\%} &   49.9\% & \textbf{57.3\%} & 52.4\% & \textbf{55.9\%}\\ 
 M3 AR & 50.8\%  & \textbf{55.5\%} & 51.3\% & \textbf{60.1\%}& 53.8\% & \textbf{56.8\%}\\ 
 CVaR AR & 63.8\% & \textbf{73.5\%} & 72.3\% & \textbf{84.3\%} & 75.0\% & \textbf{76.1\%}\\ \hline 
 Raw Mixer Layer Duration & 320dt  & 320dt & 320dt & 320dt& 320dt& 320dt \\ 
 PO Mixer Layer Duration & -  & \textbf{128dt}  & -  & \textbf{128dt} & - & \textbf{128dt}\\ 
 \hline
 \end{tabular}
\caption{Benchmark the hybrid gate-pulse model for QAOA Max-Cut problem on a three-regular six nodes graph. AR refers to the Approximate Ratio. GO is the gate-level optimization. M3 is a measurement error mitigation technique. And PO refers to pulse-level optimization.}
\label{table:qaoa-Max-Cut}
\vspace{-4mm}
\end{table*} 

\textbf{Gate-level Optimization and Error Mitigation:} In Table.~\ref{table:qaoa-Max-Cut}, we compare the performance between the hybrid model and standard gate-level QAOA with experimental results collected from real NISQ machines. We apply gate-level optimization methods, including Sabre mapping and commutative gate cancellation, on both models. Then, we implement the M3. From the results on $ibmq\_toronto$, we validate that both of the gate-level optimization methods and measurement error mitigation are compatible with our proposed model, achieving 3.2\% and 2.8\% improvement towards the approximation ratio, respectively. 
Moreover, our proposed hybrid model consistently outperforms the gate-level model. 
For the raw hybrid gate-pulse model, we have the advantage of 5.3\% in the approximation ratio compared to the raw gate-level model. With the gate-level optimization methods, our model is 7.4\% more accurate in approximation ratio than the gate-level model. With M3 implemented on both models, our proposed model has a 60.1\% approximation ratio, which is 8.8\% higher than the gate-level model obtained.

 \begin{figure}[t]
\centering
\includegraphics[width=1\linewidth]{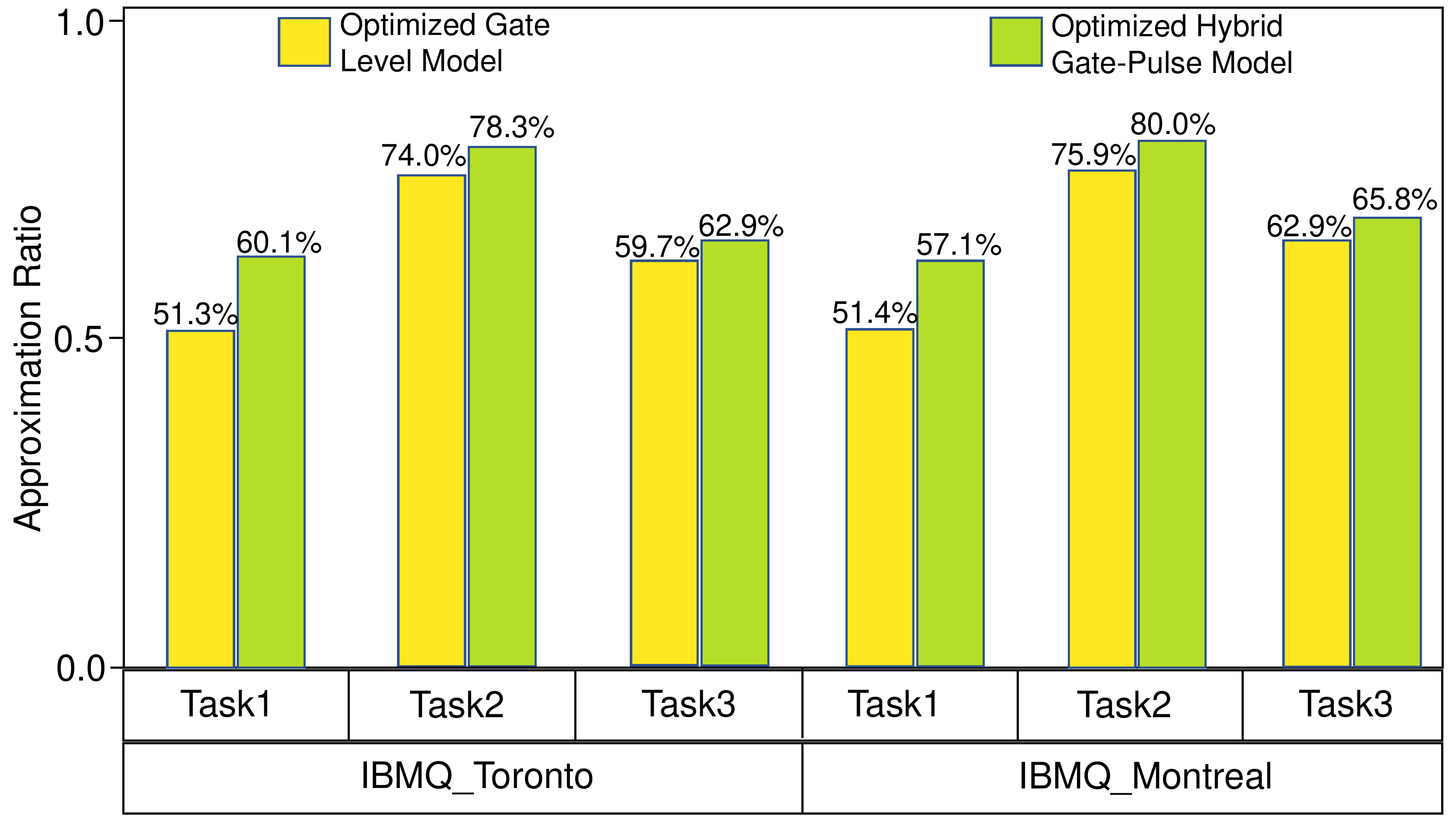}
\caption{Optimized Gate-level model and optimized hybrid gate-pulse model for QAOA Max-Cut problem with three-regular six nodes, randomized six nodes, and three-regular eight nodes graphs
on $ibmq\_toronto$ and $ibmq\_montreal$.}
\label{fig:threegraph}
\vspace{-5mm}
\end{figure}

\textbf{More NISQ Machine Results: }We evaluate the hybrid model on three different NISQ machines provided by IBM with a three-regular six nodes graph. Table. \ref{table:qaoa-Max-Cut} depicts the ability of the proposed model over the gate-level model. For raw models, with gate-level optimization, with M3, and with CVaR, the hybrid gate-pulse model demonstrates improvements in approximation ratio by 4.7\%, 4.4\%, and 5.5\% compared to on gate-level model on the average of results obtained from NISQ machines. Pulse-level optimization generally reduces duration by 60\% on the mixer layer of QAOA. We further evaluate both the optimized gate-level model and the optimized hybrid gate-pulse model; for both models, gate-level optimization and M3 have been applied, and the hybrid gate-pulse model adds the pulse-level optimization. The results indicate proposed model achieves on average 7.3\%, 4.2\%, 3.0\% higher than the gate-level model on task 1, task 2, task 3, respectively.


 \section{Conclusion and Open Questions}
 \label{sec:conclusion}
 In this work, we propose the first hybrid gate-pulse model as a new abstraction layer to design and implement variational quantum algorithms. With experimental results from IBM's near-term quantum processors, we demonstrate performance improvements on the Max-Cut problem using a hybrid QAOA to reduce pulse duration and enhance the approximation ratio. Additionally, our framework integrates multiple general optimization techniques such as measurement reduction, Conditional Value-at-Risk, and measurement error mitigation. Finally, we reach a 8\% performance boost and 60\% duration reduction on the mixer layer compared to existing methods. We also list many other optimization methods that can be adopted into our hybrid model for future work. Our hybrid model can be naturally adopted into many other VQAs for Quantum Machine Learning (QML) purposes such as quantum generative models~\cite{zoufal2019quantum} and quantum classifiers~\cite{schuld2020circuit, li2021vsql}. One counterexample is the quantum autoencoder for data compression~\cite{romero2017quantum} since pulse-level contains more parameters and hence reduces the compression ratio.
 
 Compared to current gate-level VQA designs, we see the potential of using pulse-level abstraction in the hybrid model to construct more efficient problem-specific Hamiltonian layers in terms of parameterized multi-quit gates. Relevant works have the potential to reduce redundancy introduced when encoding the problem. Notably, in our work, we define the problem-agnostic part as our pulse layer in the hybrid gate-pulse abstraction layer. However, we believe the choice of how to use the pulse layer and gate layer in the hybrid gate-pulse abstraction layer for specific problems can still be an open question. For example, if we adopt the idea of the hybrid gate-pulse model to quantum neural network tasks, the pulse-level encoding layer may benefit more since the pulse has more parameters that may have the potential to allow more classical data encode to the quantum state, and the gate-level trainable layer has lots of existing techniques that could be used to gain the advantage. On the other hand, the general trainability and further optimization of the pulse-level abstraction layer is still an open question since it involves a larger parameter space and may lead to problems such as Barren Plateaus~\cite{mcclean2018barren}.
 
\section{Acknowledgement}
\label{sec:acknowledgement}
The authors would like to thank Naoki Kanazawa for insightful discussions on pulse-level algorithm implementation and design through Qiskit-OpenPulse, Jiaqi Leng and Yuxiang Peng for discussion on pulse-level gradient descent methods. We acknowledge the use of IBM Quantum services for this
work. 
\printbibliography
\end{document}